\documentclass[11pt]{article}
\usepackage{hyperref}
\pdfoutput=1
\begin{document}
\title{Bubble bean bags in shampoo}
\author{Anup Kundu, Chandan Sharma, Gargi Das, G Harikrishnan\\
\\\vspace{6pt} Department of Chemical 
Engineering, \\ Indian Institute of Technology Kharagpur, West Bengal, 721302, India}
\maketitle
\begin{abstract}
In these fluid dynamics videos, we, for the first time, show various interactions of a 'Taylor bubble' with their smaller and differently shaped counterparts, in a shear thinning, non-Newtonian fluid, confined in a narrow channel. 
\end{abstract}
% main text

The dynamics of bubbles in a quiescent liquid is an important problem in multiphase flow \textsuperscript{1, 2}. Small volumes of dispersed air introduced into a liquid confined in a conduit generates spherical or cap shaped bubbles, whereas air volumes larger than a critical value give rise to elongated, axisymmetric and bullet shaped 'Taylor bubbles'. In this fluid dynamics video, we demonstrate various interactions of Taylor bubbles with their smaller counterparts. The medium in which the interactions are recorded is a common non-Newtonian (shear thinning) liquid (shampoo).

A glass tube of inner diameter 12 mm and length 1.5 m is completely filled with a commercial shampoo (Pantene Pro V). Small volumes of air are introduced at the bottom of the tube, using a syringe. After the small bubbles rise to a certain height in the channel, a bigger Taylor bubble is introduced at the bottom. The faster rising Taylor bubble interacts with the slowly upward moving satellite bubbles. The videos of the interactions are recorded by a digital SLR camera, mounted on a tripod.

We observe that a comparatively larger spherical bubble gets deformed and sits on the tip of the Taylor bubble. Both bubbles then traverse upward as a single unit. A relatively smaller bubble slips from the tip and gets entrapped in the draining liquid. This bubble is pushed downward through the annular liquid film draining between the elongated bubble and the tube wall. During its downward motion, the smaller bubble pushes the gas - liquid interface of the Taylor bubble inward. The deformation of the interface is reversible. After the passage of the small bubble the Taylor bubble regains its original shape, resembling a cushion or a bean bag. We speculate that this is due to the viscoelastic nature of the air-liquid interface. The small bubble comes out of the draining liquid and moves into the bulk fluid, while the Taylor bubble moves up. 
    
A unique phenomenon occurs when two equal sized bubbles interact with the Taylor bubble. Initially, both the bubbles sit one above the other on the Taylor bubble tip. After a while, the middle bubble is drained along the liquid film while the upper one sits on the tip. The falling bubble deforms the interface as described above. However when it comes out of the draining liquid, it changes its direction of motion and gets attached to the tail of the Taylor bubble. At this point, the motion of the small bubble shows a component of velocity perpendicular to the direction of motion of the Taylor bubble. The Taylor bubble now translates upward with satellite bubbles perched at its front and rear ends. During the interactions, we did not observe coalescence of bubbles.        
%% The format is: \href{URL of video}{name that will appear in the text}

\href{http://ecommons.library.cornell.edu/bitstream/1813/8237/2/shampoo-high-res.mpg}{Video
1} and
\href{http://ecommons.library.cornell.edu/bitstream/1813/8237/4/shampoo-low-res.mpg}{Video
2}.
\begin{enumerate}
\item
R. G. Sousa, A. M. F. R. Pinto, J. B. L. M. Campos. "Interaction between Taylor bubbles rising in stagnant non-Newtonian fluids." Int. J. Multiphase Flow, \textbf{33}, 970 (2007). 

\item
T. K. Mondal, G. Das and P. K. Das. "Prediction of rise velocity of a liquid Taylor bubble in vertical tube," Phys. Fluids, \textbf{19}, 128109 (2007).
\end{enumerate}
\end{document}